\documentclass[prl,twocolumn,tightenlines,aps,epsf,showpacs,superscriptaddress,longbibliography,amsmath,amssymb]{revtex4-2}

\usepackage[pdftex]{graphicx}
\usepackage{dcolumn}
\usepackage{bm}
\usepackage{epsfig}
\usepackage{latexsym} 
\usepackage{amsmath}
\usepackage{amssymb}
\usepackage{color}
\usepackage{array}
\usepackage{bbm}
\usepackage{color}
\usepackage{array}
\usepackage{cancel}
\usepackage[dvipsnames]{xcolor}
\usepackage{float}
\usepackage[normalem]{ulem}
\usepackage[mathscr]{euscript}
\usepackage{grffile}
\usepackage{physics}
\usepackage[hidelinks]{hyperref}%
\usepackage{titlesec}
\titleformat{\paragraph}[runin]
        {\bfseries}
        {}
        {0.0em}
        {}
        [ -- ~]
\titlespacing*{\paragraph}{0pt}{4pt}{0pt}

\usepackage[normalem]{ulem} 

\usepackage{amsmath} 

\makeatletter
\@addtoreset{equation}{section}
\@addtoreset{figure}{section}
\makeatother
\usepackage{tikz}
\begin{document}

\title{A Nanomechanical Atomic Force Qubit}

\author{Shahin Jahanbani}
\affiliation{Department of Physics, University of California, Berkeley, California 94720, USA}
\affiliation{Materials Sciences Division, Lawrence Berkeley National Laboratory, Berkeley, California 94720, USA}

\author{Zi-Huai Zhang}
\affiliation{Department of Electrical Engineering and Computer Sciences,
University of California, Berkeley, California 94720, USA}
\affiliation{Materials Sciences Division, Lawrence Berkeley National Laboratory, Berkeley, California 94720, USA}
\affiliation{Department of Physics, University of California, Berkeley, California 94720, USA}

\author{Binhan Hua}
\affiliation{Quantum Science and Engineering, Harvard University, Cambridge, MA 02138}

\author{Kadircan Godeneli}
\affiliation{Department of Electrical Engineering and Computer Sciences,
University of California, Berkeley, California 94720, USA}
\affiliation{Materials Sciences Division, Lawrence Berkeley National Laboratory, Berkeley, California 94720, USA}

\author{Boris Müllendorff}
\affiliation{Department of Electrical Engineering and Computer Sciences,
University of California, Berkeley, California 94720, USA}

\author{Xueyue Zhang}
\affiliation{Department of Electrical Engineering and Computer Sciences,
University of California, Berkeley, California 94720, USA}
\affiliation{Department of Physics, University of California, Berkeley, California 94720, USA}

\author{Haoxin Zhou}
\affiliation{Department of Electrical Engineering and Computer Sciences,
University of California, Berkeley, California 94720, USA}
\affiliation{Materials Sciences Division, Lawrence Berkeley National Laboratory, Berkeley, California 94720, USA}
\affiliation{Department of Physics, University of California, Berkeley, California 94720, USA}

\author{Alp Sipahigil}
\email{alp@berkeley.edu}
\affiliation{Department of Electrical Engineering and Computer Sciences,
University of California, Berkeley, California 94720, USA}
\affiliation{Materials Sciences Division, Lawrence Berkeley National Laboratory, Berkeley, California 94720, USA}
\affiliation{Department of Physics, University of California, Berkeley, California 94720, USA}
\date{\today}

\begin{abstract}
Silicon nanomechanical resonators display ultra-long lifetimes at cryogenic temperatures and microwave frequencies. Achieving quantum control of single-phonons in these devices has so far relied on nonlinearities enabled by coupling to ancillary qubits. In this work, we propose using atomic forces to realize a silicon nanomechanical qubit without coupling to an ancillary qubit. The proposed qubit operates at $60$~MHz with a single-phonon level anharmonicity of 5 MHz. We present a circuit quantum acoustodynamics architecture where electromechanical resonators enable dispersive state readout and multi-qubit operations. The combination of strong anharmonicity, ultrahigh mechanical quality factors, and small footprints achievable in this platform could enable quantum-nonlinear phononics for quantum information processing and transduction.  
\end{abstract}
\maketitle

\indent Engineered quantum systems based on superconducting  \cite{blais2004cavity,blais2007quantum,blais2020quantum}, mechanical \cite{kalaee2019quantum,fink2016quantum,chu2018creation} and photonic circuits \cite{sipahigil2016integrated,kindem2020control,thompson2013coupling} require strong, single-quantum level nonlinearities for quantum state control and readout. Such strong nonlinearities are routinely achieved with Josephson junctions in superconducting quantum processors \cite{blais2021circuit}. Despite becoming a leading platform for quantum computing, superconducting circuits still face open challenges related to their large footprints and short lifetimes \cite{wang2015surface,martinis2022surface}. In comparison, silicon nanomechanical resonators display ultra-long lifetimes and small footprints \cite{maccabe2020nano} but lack the nonlinearities needed for quantum-level control and readout of single phonons. Recent efforts have developed mechanical modes coupled to ancillary superconducting qubits~\cite{wollack2022quantum,manenti2017circuit,o2010quantum,yang2024} to achieve quantum-level control and readout of single phonons. However, these hybrid systems suffer from the coherence limitations of ancillary qubits and heterogeneous integration challenges. In this work, we propose the realization of a nanomechanical qubit without the need for coupling to an ancillary qubit. In this platform, a strong single-phonon nonlinearity is achieved with the surface atomic forces between a cantilever oscillator and a tip.
We describe a circuit quantum acoustodynamics (cQAD) architecture that integrates the nanomechanical qubit with electromechanical and microwave resonators to achieve dispersive state readout and multi-qubit operations.\\
\begin{figure}[hbt!]
    \centering
    \includegraphics[width=1\linewidth]{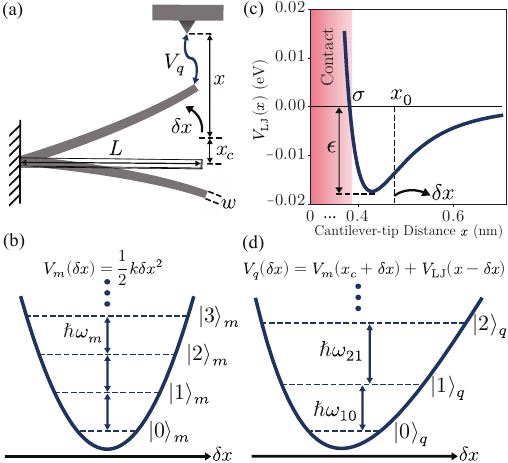}
    \caption{\textbf{A nanomechanical qubit based on atomic forces}. \textbf{(a)} Cantilever has length $L$, width $w$, thickness $t$ (out of plane), oscillation amplitude $\delta x$. Cantilever is biased at equilibrium position $x_c$ and cantilever-tip distance $x$. \textbf{(b)} At the single-phonon regime, the isolated cantilever is a quantum harmonic oscillator. \textbf{(c)} Lennard-Jones potential, $V_{\text{LJ}}$, as a function of cantilever-tip distance $x$. The approximate contact region is shaded in red. The nanomechanical qubit is biased using a MEMS actuator at the position $x=x_0\approx1.24\sigma$, where the spring constant induced by the Lennard-Jones potential vanishes.  The parameters of the shown potential are for the silicon-silicon interface with $\epsilon = 17.4\ \text{meV}$ and $\sigma = 3.826\ \text{Å}$~\cite{inui2017interaction}. \textbf{(d)} The combined potential results in an anharmonic quantum oscillator that operates as a nanomechanical qubit at frequency $\omega_{10}$.}
    \label{Fig. 1}
\end{figure}
\indent Our proposal for the nanomechanical qubit builds on the principles of atomic force microscopy (AFM)~\cite{binnig1986atomic}. In AFM, material surface topography is probed via the deflections of a cantilever tip brought near a surface. The spatial resolution originates from distance-sensitive atomic forces between the tip and the surface that perturb the cantilever. Assuming a pairwise atom-atom tip-surface interaction, the surface atomic forces can be modeled by the Lennard-Jones potential, $V_{\text{LJ}}\left(x\right)=4\epsilon\left[\left(\sigma/x\right)^{12}-\left(\sigma/x\right)^6\right]$, where $x$ is the tip-surface static distance, $\epsilon$ is the potential depth, and $\sigma$ is the potential offset distance \cite{gould1989simple,chen2021introduction,giessibl2003advances,Ashcroft:102652,giessibl2003advances}. The parameters $\epsilon$ and $\sigma$ are determined by material properties. We use a silicon-based design, and the parameters used are for the silicon-silicon interface. 

The main idea of the proposed nanomechanical qubit is illustrated in Fig.~\ref{Fig. 1}(a). An isolated cantilever is a harmonic oscillator with potential $V_m(\delta x)=k\delta x^2/2$ (Fig.~\ref{Fig. 1}(b)), where $\delta x$ is the oscillator amplitude. When the cantilever is brought close to a tip, the atomic forces ($V_{\text{LJ}}$, Fig.~\ref{Fig. 1}(c)) between the cantilever and the tip surface influence the harmonic potential. The addition of the Lennard-Jones potential turns the cantilever into a quantum anharmonic oscillator with the effective potential $V_q(\delta x) = V_m(x_c+\delta x)+V_{\text{LJ}}(x-\delta x)$, where $x$ is the static distance between the cantilever and the tip, and $x_c$ is the equilibrium position of the cantilever. The equilibrium position $x_c$ is determined by balancing the static force from the Lennard-Jones potential and the restoring cantilever spring force. A voltage-controlled microelectromechanical (MEMS) actuator can be used to tune the tip position and the cantilever-tip distance $x$. The resulting nonlinear potential $V_q(\delta x)$ (Fig.~\ref{Fig. 1}(d)) enables the operation of the cantilever as a nanomechanical qubit. In what follows, we optimize this system to achieve strong single-phonon level nonlinearities and high-fidelity qubit operations. \\
\indent The fidelity of qubit operations depends on the interplay between the qubit anharmonicity ($\eta=\omega_{21}-\omega_{10}$) \cite{koch2007charge} and phonon dissipation rate ($\gamma$). Assuming a frequency-independent mechanical quality factor ($Q=\omega_{10}/\gamma$), the figure of merit for high-fidelity operations is characterized by the relative anharmonicity, which is the ratio of the anharmonicity and the qubit frequency ($\eta_r\equiv\eta/\omega_{10}$) \cite{koch2007charge,pistolesi2021proposal,hyyppa2022unimon}. The relative anharmonicity for the nanomechanical qubit is 
\begin{align}
\eta_r\approx\frac{1+2r_1(x)}{1+r_1(x)+r_0(x)},
\label{Eq. 1}
\end{align}
where $r_0(x)=2\hbar\omega_{\text{eff}}/(x_{\text{zpf}}^4V_{\text{LJ}}^{(4)}(x))$ and $r_1(x)=V_{\text{LJ}}^{(6)}(x)x_{\text{zpf}}^2/(4V_{\text{LJ}}^{(4)}(x))$. Here, $V_{\text{LJ}}^{(n)}\left(x\right)$ is the $n$-th derivative of the Lennard-Jones potential, $\omega_{\text{eff}}=\sqrt{(k+k_{\text{LJ}})/m_{\text{eff}}}$ is the effective cantilever angular frequency with the effective mass $m_{\text{eff}}\approx0.24\rho L w t$, native spring constant $k=Etw^3/(4L^3)$, and $k_{\text{LJ}}=|V_{\text{LJ}}^{\left(2\right)}|$, which is the spring constant induced by the Lennard-Jones potential, and  $x_{\text{zpf}}=\sqrt{\hbar/(2m_{\text{eff}}\omega_{\text{eff}})}$ is the zero-point motion of the cantilever lateral oscillation (see Supplemental Material Sec.~II \cite{Supp}). $(L,w,t)$ are the cantilever dimensions, $E$ is Young's modulus, and $\rho$ is the volume mass density. We maximize $\eta_r$ by optimizing the nanomechanical oscillator geometry. We use $E = 160\ \text{GPa}$ and $\rho = 2329\ \text{kg/m}^3$ for silicon \cite{hopcroft2010young,maluf2004introduction,henins1964precision,fujii1995absolute} and reduce the optimization parameter space for $\eta_r$ in Eq. \ref{Eq. 1} by choosing small transverse dimensions $(w,t)=(10,12)\ \text{nm}$, which produce a large zero-point motion and anharmonicity. These dimensions are routinely achieved in silicon nanodevices such as fin field-effect transistors \cite{ChenmingHu2000}.\\
\begin{figure}[t]
    \centering
    \includegraphics[width=1\linewidth]{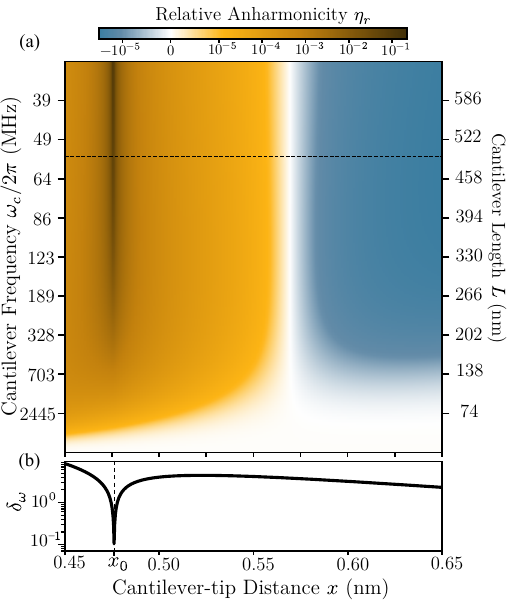}
    \caption{\textbf{Quantum anharmonicity of the nanomechanical qubit}. \textbf{(a)} Relative anharmonicity $\eta_r$  as a function of the cantilever-tip distance $x$, cantilever length $L$, and isolated cantilever frequency $\omega_c\propto L^{-2}$. \textbf{(b)} The relative frequency shift $\delta_\omega$ as a function of the cantilever-tip distance at the isolated cantilever frequency $\omega_c=2\pi\times55\ \text{MHz}$ (dashed line in (a)). We choose to bias the qubit at $x_0\approx1.24\sigma$ where the relative frequency shift is at its minimum, and the relative anharmonicity is maximized. At $x=x_0$, $\eta_r\approx0.089$ is achieved for the cantilever dimensions $(L,w,t) = (495,10,12)\ \text{nm}$ and the qubit frequency $\omega_q=2\pi\times60\ \text{MHz}$.}
    \label{Fig. 2}
\end{figure} 
\begin{figure*}[htbp]
    \centering
    \includegraphics[width=1\linewidth]{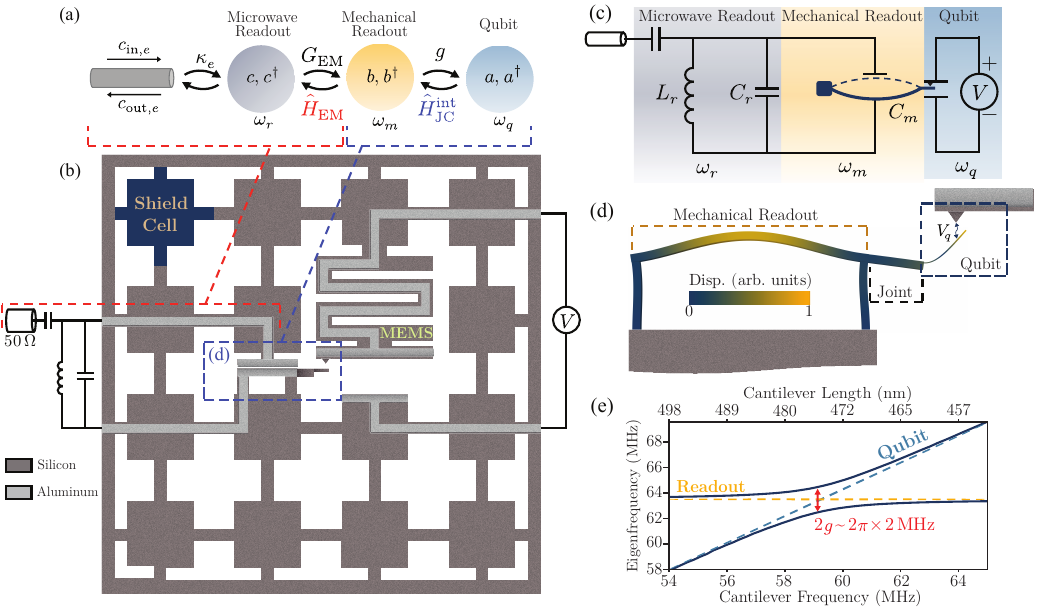}
    \caption{\textbf{Circuit quantum acoustodynamics setup.} \textbf{(a)} A mechanical readout resonator $(\omega_m)$ is coupled to the nanomechanical qubit ($\omega_q$) and microwave resonator $(\omega_r)$ modes with respective coupling rates $g$ and $G_{\text{EM}}$. The external coupling rate to the microwave transmission line is $\kappa_e$, and $c_{\text{in},e} (c_{\text{out},e})$ denotes input (output) fields. \textbf{(b)} The quantum electromechanical system containing the nanomechanical qubit, mechanical, and microwave readout resonators. The nanomechanical qubit is realized as a silicon nanowire cantilever. The cantilever-tip distance is tuned using a MEMS actuator. The nanomechanical qubit is hybridized with a mechanical readout resonator. The proposed platform is patterned on a suspended SOI membrane with a 220 nm thick device layer and embedded in a phononic crystal shield to eliminate phonon radiation \cite{kalaee2019quantum}. The thickness of the aluminum electrodes is 60 nm. \textbf{(c)} Circuit diagram of the cQAD setup. The mechanical readout resonator is probed with a microwave resonator with inductance $L_r$ and capacitance $C_r$. $C_m$ denotes the vacuum gap capacitance modulated by the mechanical mode. \textbf{(d)} Displacement mode profile of the hybridized nanomechanical qubit and readout resonator system. \textbf{(e)} Finite element modeling of the qubit-readout mechanical coupling strength $g$. For the joint length of $1.25~\mu$m used here, $\omega_m$ is shifted by $\sim 3\ \text{MHz}$. The mechanical readout resonator models include a 60 nm thick aluminum layer on silicon. See Supplemental Material Sec. I \cite{Supp} for the band structure and dimensions of the phononic shield cell and the mechanical resonator.} 
    \label{Fig. 3}
\end{figure*}
Bringing the cantilever near the tip results in frequency shifts due to $k_\textrm{LJ}$. We define the relative frequency shift, which quantifies the difference between the isolated cantilever and nanomechanical qubit frequencies, as $\delta_\omega=\left|1-\omega_{10}/\omega_c\right|$, where $\omega_c\approx\sqrt{Ew^2/(\rho L^4)}$ is the isolated cantilever angular frequency. Fig. \ref{Fig. 2} shows the relative anharmonicity and relative frequency shift ($\delta_\omega$) as functions of cantilever-tip distance $x$ and cantilever length $L$. We choose to operate the qubit at a distance where $\delta_\omega$ is at its minimum. In this regime, with $k_{\text{LJ}}=0$, the surface forces contribute a static force (balanced by the MEMS actuator) and a nonlinear restoring force that provides a Kerr-type, third-order mechanical nonlinearity. As a result, we have $\omega_{\text{eff}}\approx \omega_c$ and the anharmonicity $\eta$, which roughly scales as $\omega_{\text{eff}}^{-2}$, is maximized (see Supplemental Material Sec.~I~\cite{Supp}). This condition is met at $x_0\approx1.24\sigma$ (Fig.~\ref{Fig. 2}(b)) for the specific Lennard-Jones potential parameters. 

Biasing the qubit to $x=x_0$ allows us to determine the qubit frequency using cantilever dimensions alone and leaves the cantilever length $L$ as the only remaining parameter in the qubit design. Cantilever length $L$ is determined by a trade-off between anharmonicity and mean thermal occupancy (${\bar{n}}_{\text{th}}\approx1/\left(e^{\hbar\omega_{10}/k_BT}-1\right)$), which both increase with increasing $L$. We choose an operation point for $L$ corresponding to sufficiently large anharmonicity while keeping thermal occupancy as low as possible. We choose $L=495\ \text{nm}$ corresponding to the nanomechanical qubit frequency of $\omega_q=\omega_{10}=2\pi\times60\ \text{MHz}$ for $\omega_c=2\pi\times55\ \text{MHz}$, $x_{\text{zpf}}=2.14\ \text{pm}$, relative anharmonicity $\eta_r = 8.9\%$, and a mean thermal occupancy of ${\bar{n}}_{\text{th}}\approx2.3$ at $T = 8\ \text{mK}$. We discuss initialization to ${\bar{n}}_{\text{th}}\ll1$ via electromechanical cooling in the following section. A second set of parameters that achieves ${\bar{n}}_{\text{th}}\ll1$ with weaker anharmonicities are presented in Supplemental Material Sec.~I \cite{Supp}. For the above-mentioned parameters, the anharmonicity $\eta/2\pi=5\ \text{MHz}$ is about nine orders of magnitude larger than the dissipation rates $\gamma/2\pi\sim 6\ \text{mHz}$ estimated based on the mechanical quality factors of $Q\sim 10^{10}$ recently achieved in silicon phononic crystal resonators \cite{maccabe2020nano}. The large ratio of anharmonicity to dissipation rate enables high-fidelity operations for the nanomechanical qubit.   

\paragraph{Circuit quantum acoustodynamics architecture} The nanomechanical qubit Hamiltonian can be approximated as ${\hat{H}}_q/\hbar\approx(1-\eta_r/2)\omega_qa^\dag a+\eta_r\omega_q(a^\dag)^2a^2/2$. Phononic circuit engineering and the similarity of this Hamiltonian with superconducting transmon qubits \cite{koch2007charge} allow circuit quantum electrodynamics (cQED) ideas to be  implemented in this platform (see Supplemental Material Sec.~III~\cite{Supp}). In the following, we propose a cQAD architecture for the dispersive readout and long-range coupling of nanomechanical qubits. This architecture consists of a mechanical readout mode with resonance $\omega_m$ coupled to the nanomechanical qubit with resonance $\omega_q$, constituting a phononic Jaynes–Cummings-type interaction with the interaction Hamiltonian ${\hat{H}}_{\text{JC}}^{\text{int}}=\hbar g(a^\dag b+ab^\dag)$, where $a$ ($b$) is the annihilation operator for the nanomechanical qubit (nanomechanical readout resonator), and $g$ is the single-phonon coupling strength (see Fig. \ref{Fig. 3}(a)). This interaction can be physically realized using the platform shown in Fig.~\ref{Fig. 3}(b). The proposed device is patterned on a suspended silicon-on-insulator (SOI) membrane with a silicon device layer thickness of 220 nm \cite{mirhosseini2020superconducting}. The nanomechanical qubit is realized by bringing the cantilever near a tip. The cantilever-tip distance can be tuned using a MEMS actuator that adjusts the strength of the nonlinear restoring forces. Unlike in AFM, the lithographically defined cantilever and tip constitute a monolithic device. The cantilever-tip distance will be robust against environmental vibrations that cause common mode displacement. The cantilever is partially etched to the desired thickness and patterned at the end of a double-sided-clamped flexural mechanical resonator at frequency $\omega_m=2\pi\times67\ \text{MHz}$. The entire structure is embedded within a phononic crystal shield to eliminate phonon radiation (Fig.~\ref{Fig. 3}(b)). The flexural mode and the nanobeam lateral mode are in the same direction, enabling strong qubit-cavity coupling. The coupling strength can be engineered by adjusting an extension joint  (labeled in Fig.~\ref{Fig. 3}(d)). Note that adding such a joint results in a slight shift from the target value $\omega_m = 2\pi\times67\ \text{MHz}$ (see Fig.~\ref{Fig. 3}(e) and Supplemental Material Sec.~I~\cite{Supp}). In this design, the qubit-resonator mode hybridization is optimized for a coupling strength of $g=2\pi\times1\ \text{MHz}$ (Fig.~\ref{Fig. 3}(e)). The system operates in the dispersive limit ($\Delta = \omega_m-\omega_q\gg g$) in which the qubit-cavity coupling enables the qubit-state-dependent shift of the mechanical resonator resonance ($\omega_m\rightarrow\omega_m\pm\chi$), where $\chi=-g^2\eta_r\omega_q/(\Delta(\Delta+\eta_r\omega_q))$ \cite{blais2004cavity,schuster2005ac,blais2021circuit,blais2020quantum,wallraff2004strong}. The dispersive shift of $\chi= -2\pi\times160\ \text{kHz}$ places the system in the strong dispersive regime even for modest mechanical quality factors of  $Q\sim 10^{4}$.\\
\indent We use cavity electromechanical interactions to cool and measure the mechanical Jaynes-Cummings system using microwave fields (see Fig.~\ref{Fig. 3}(a)). A microwave resonator (labeled with resonance frequency $\omega_r$ in Fig.~\ref{Fig. 3}(a)) is shunted by a vacuum-gap capacitor ($C_m$) on the mechanical readout resonator. Mechanical vibrations modulate the capacitance and result in electromechanical coupling \cite{kalaee2019quantum,fink2016quantum} (see Supplemental Material Sec. III \cite{Supp}). Under a red-detuned microwave drive field ($\omega_d \approx \omega_r-\omega_m$), the microwave and mechanical modes are parametrically coupled with a beamsplitter Hamiltonian ${\hat{H}}_{\text{EM}}=\hbar G_{\text{EM}}(b^\dag c+bc^\dag)$, where $c$ denotes the microwave field and $G_{\text{EM}}$ is the parametric coupling strength that depends on the drive power \cite{aspelmeyer2014cavity}. For $G_{\text{EM}}\ll\omega_m,\kappa$, the mechanical modes are coupled to the microwave input line with an effective electromechanical damping rate $\Gamma_{e}\approx4 G_{\text{EM}}^2/\kappa$, where $\kappa$ denotes the total microwave resonator damping rate (Supplemental Material Sec. III \cite{Supp}). This electromechanical damping allows both the sideband cooling of the mechanical mode occupancy to ${\bar{n}}_{\text{th}}\ll1$ \cite{teufel2011sideband,chan2011laser} and the interrogation of the mechanical system using microwaves. \\
 \begin{figure}[t]
    \centering   
    \includegraphics[width=1\linewidth]{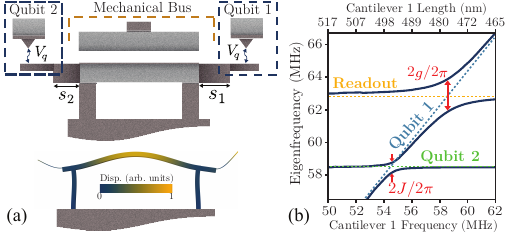}
    \caption{\textbf{Bus-mediated qubit-qubit coupling} \textbf{(a)} The mechanical readout resonator is dispersively coupled to two nanomechanical qubits. The qubit-qubit coupling is mediated via virtual phonon exchanges with the cavity, where the coupling strengths can be tuned by adjusting lengths $s_1$ and $s_2$. \textbf{(b)} Finite element modeling of the qubit-resonator and qubit-qubit interactions strengths extracted from avoided crossings of respective strengths $g$ and $J$.}
    \label{Fig. 4}
\end{figure}
\indent As an example of the scalability of the cQAD architecture, we discuss the implementation of two-qubit interactions via a mechanical bus \cite{majer2007coupling,sillanpaa2007coherent}. The design shown in Fig.~\ref{Fig. 4}(a) attaches two nanomechanical qubits to the structure shown in Fig.~\ref{Fig. 3}(d). In this structure, the mechanical resonator is dispersively coupled to each qubit and mediates the interactions. Each qubit virtually exchanges phonons with the cavity bus, implementing indirect qubit-qubit coupling with strength $J$~\cite{majer2007coupling}. Fig.~\ref{Fig. 4}(b) shows the finite element modeling of the system, revealing two avoided crossings corresponding to qubit-cavity and qubit-qubit mechanical mode hybridizations. Based on the simulation, we find $J\approx2\pi\times0.3\ \text{MHz}$, comparable to the theoretically estimated value of $J\approx2\pi\times0.23\ \text{MHz}$, assuming degenerate qubit modes with identical qubit-cavity coupling strengths (Supplemental Material Sec.~III \cite{Supp}). We note that longer-range, microwave-mediated couplings could also be achieved using the electromechanical beamsplitter interactions discussed above.

\indent In summary, we introduced the idea of using surface atomic forces to realize a nanomechanical qubit. We described a cQAD architecture where the nanomechanical qubit is implemented by a cantilever, and read out via mechanical and microwave resonators. We also discussed an example implementation of two-qubit interactions via a mechanical bus. The proposed cQAD implementation focuses on a design that maximizes qubit anharmonicity using a low-frequency cantilever with a large zero-point motion. The required dimensions, while regularly achieved in state-of-the-art semiconductor foundries, pose experimental challenges for initial experiments. Early experiments might benefit from designs with higher stiffness and frequencies at the cost of reduced anharmonicities (parameters discussed in Supplemental Material Sec.~I \cite{Supp}). We also point that surface details of the experimental devices can cause deviations from the Lennard-Jones potential $V_{\text{LJ}}$ used here. The biasing strategies and results discussed here are applicable to modified nonlinear potentials, which will change the qubit nonlinearity. Achieving stronger nonlinear surface forces while using fabrication-friendly parameters might be possible by using different materials, electrostatic or magnetostatic interactions. The combination of strong anharmonicity, ultrahigh mechanical quality factors, and small footprints for the proposed platform might enable future quantum sensing, transduction, and computing technologies based on purely mechanical quantum systems. 

\vspace{4pt}\noindent{\it Acknowledgements} -- We thank Feng Wang, Michael Crommie, and Daniel Carney for valuable discussions. This work was primarily funded by the U.S. Department of Energy, Office of Science, Basic Energy Sciences, Materials Sciences, and Engineering Division under Contract No. DE-AC02-05CH11231 within the Nanomachine Program. Finite element modeling of the proposed devices was supported by the U.S. Department of Energy, Office of Science, Office of Basic Energy Sciences, Materials Sciences and Engineering Division under Contract No. DE-AC02-05-CH11231 in the Phonon Control for Next-Generation Superconducting Systems and Sensors FWP (KCAS23). 

\bibliography{main}
\clearpage

\end{document}


\title{Supplemental Material for \\``A Nanomechanical Atomic Force Qubit''}

\author{Shahin Jahanbani}
\affiliation{Department of Physics, University of California, Berkeley, California 94720, USA}
\affiliation{Materials Sciences Division, Lawrence Berkeley National Laboratory, Berkeley, California 94720, USA}

\author{Zi-Huai Zhang}
\affiliation{Department of Electrical Engineering and Computer Sciences,
University of California, Berkeley, California 94720, USA}
\affiliation{Materials Sciences Division, Lawrence Berkeley National Laboratory, Berkeley, California 94720, USA}
\affiliation{Department of Physics, University of California, Berkeley, California 94720, USA}

\author{Binhan Hua}
\affiliation{Quantum Science and Engineering, Harvard University, Cambridge, MA 02138}

\author{Kadircan Godeneli}
\affiliation{Department of Electrical Engineering and Computer Sciences,
University of California, Berkeley, California 94720, USA}
\affiliation{Materials Sciences Division, Lawrence Berkeley National Laboratory, Berkeley, California 94720, USA}

\author{Boris Müllendorff}
\affiliation{Department of Electrical Engineering and Computer Sciences,
University of California, Berkeley, California 94720, USA}

\author{Xueyue Zhang}
\affiliation{Department of Electrical Engineering and Computer Sciences,
University of California, Berkeley, California 94720, USA}
\affiliation{Department of Physics, University of California, Berkeley, California 94720, USA}

\author{Haoxin Zhou}
\affiliation{Department of Electrical Engineering and Computer Sciences,
University of California, Berkeley, California 94720, USA}
\affiliation{Materials Sciences Division, Lawrence Berkeley National Laboratory, Berkeley, California 94720, USA}
\affiliation{Department of Physics, University of California, Berkeley, California 94720, USA}

\author{Alp Sipahigil}
\email{alp@berkeley.edu}
\affiliation{Department of Electrical Engineering and Computer Sciences,
University of California, Berkeley, California 94720, USA}
\affiliation{Materials Sciences Division, Lawrence Berkeley National Laboratory, Berkeley, California 94720, USA}
\affiliation{Department of Physics, University of California, Berkeley, California 94720, USA}

\date{\today}
\maketitle
\label{Sec:SI}

\setcounter{figure}{0}
\setcounter{section}{0}
\renewcommand{\thefigure}{S\arabic{figure}}
\renewcommand{\thetable}{\Roman{table}}
\renewcommand{\thesection}{\Roman{section}}

\section{\label{SI_Design}I. Nanomechanical design}
\begin{figure}[hbt!]
    \centering
    \includegraphics[width=1\linewidth]{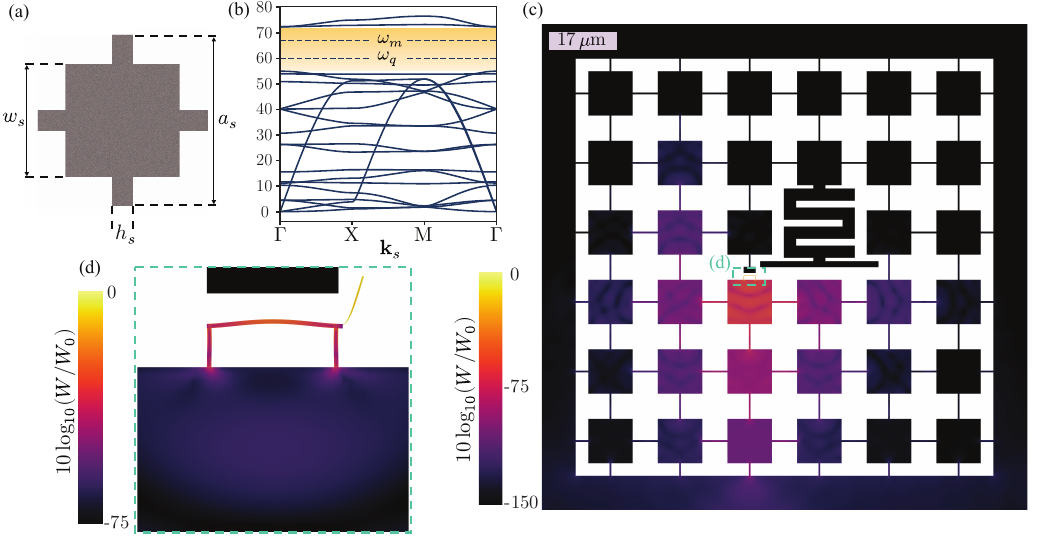}
    \caption{\textbf{Phononic crystal shield.} \textbf{(a)} Schematic of the phononic shield unit cell with dimensions $\left(a_s,w_s,h_s,t_s\right)=\left(17.25,11.01,0.44, 0.22\right)\mu \text{m}$ where $t_s$ is the shield thickness. \textbf{(b)} Mechanical band structure for the phononic crystal with a bandgap of 17.4 MHz centered at 63.7 MHz. Qubit ($\omega_q$) and readout ($\omega_m$) modes are both in the bandgap and shielded from clamping losses. \textbf{(c)} Normalized total energy density profile ($W/W_0$) of the single-qubit device at $\omega_m$ with \textbf{(d)} the zoom-in qubit-resonator region.}
    \label{SI Fig}
\end{figure}
\begin{figure}
    \centering
    \includegraphics[width=0.5\linewidth]{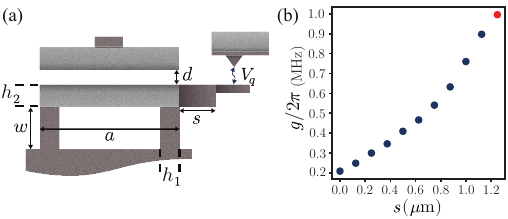}
    \caption{\textbf{Qubit-mechanical readout resonator coupling.} \textbf{(a)} Schematic of the mechanical resonator and the associated dimensions. Selection of $\left(a,w,h_1,h_2\right)=(3.031,0.9,0.1,0.1)\ \mu \text{m}$ corresponds to $\omega_m=2\pi\times67\ \text{MHz}$. The respective thicknesses of the silicon device layer and aluminum electrodes layer are 220 nm and 60 nm. {\textbf{(b)}} Simulated coupling strength $g/2\pi$ tuned via the extension joint length $s$. The desired strength is red color-coded.}
    \label{SI2 Fig}
\end{figure}
The dimensions and mechanical band structure simulation of the phononic crystal shield of the quantum device in the main text are shown in Fig.~\ref{SI Fig}(a) and Fig.~\ref{SI Fig}(b). The dimensions of the mechanical resonator and the readout coupling strength tuning are demonstrated in Fig.~\ref{SI2 Fig}. The elastic energy density profile of the quantum device is depicted in Fig.~\ref{SI Fig}(c) and Fig.~\ref{SI Fig}(d). All the mechanical simulations performed in COMSOL Multiphysics assume anisotropic single-crystal silicon with the volume mass density $\rho = 2329\ \text{kg/m}^3$, Young's modulus $E = 160\ \text{GPa}$, and elasticity matrix elements $(C_{11}, C_{12}, C_{44}) = (166, 64, 80)\ \text{GPa}$~\cite{comsol}. 

Alternative nanomechanical qubit designs can also be used to increase the operation frequency or reduce fabrication complexity at the cost of reduced nonlinearities. For $(w,t)=(10,12)~ \text{nm}$, $L\le345\ \text{nm}$, and $T=8\ \text{mK}$, we have a thermal occupancy below unity with a relative anharmonicity of $\eta_r\le2.3\%$ and higher qubit operation frequencies at ${\omega}_{10}\geq2 \pi\times115\ \text{MHz}$. For the same range of qubit frequencies, the qubit can also be initialized to ${\bar{n}}_{\text{th}}\le1$ with more fabrication-friendly parameters $\left(w,t\right)=\left(18,24\right)\ \text{nm}$ and $L\le457\ \text{nm}$. The relative anharmonicity is further reduced to $\eta_r\le0.1\%$ for these parameters.

~

\section{\label{SI_anharmonicity} II. Nanomechanical qubit anharmonicity}

In this section, we derive an expression for the nanomechanical qubit anharmonicity. Including surface atomic forces approximated by the Lennard-Jones potential ($V_{\text{LJ}}$) and adjustable electrostatic forces between the capacitor plates of the MEMS actuator that tune the strength of atomic forces, the Hamiltonian of the nanomechanical qubit can be written from Fig. 1(d) of the main text as
$H=p^2/2m_{\text{eff}}+k(x_c+\delta x)^2/2+V_{\text{LJ}}(x-\delta x)$, where $p$ is the momentum, $m_{\text{eff}}$ is the effective cantilever mass, $k$ is the native spring constant, $x_c$ is the shifted harmonic equilibrium due to static Lennard-Jones force contribution, $\delta x$ is the oscillation amplitude, and $x = x(V_{\text{MEMS}})$ is the cantilever-tip distance tuned by the DC bias voltage through MEMS actuation. Taylor expanding the Lennard-Jones potential in the powers of oscillator amplitude $\delta x$ and grouping terms, we obtain 
\begin{align}
H\approx\frac{p^2}{2m_{\text{eff}}}+\frac{1}{2}k(x_c+\delta x)^2+\sum_{n\geq0}{\lambda_n(x)(-\delta x)^n}=\frac{p^2}{2m_{\text{eff}}}+[kx_c+F_{\text{LJ}}(x)]\delta x+\frac{1}{2}k_{\text{eff}}\delta x^2+V_0(x,\delta x),
\label{Eq. A1} 
\end{align}
where $\lambda_n(x)=V_{\text{LJ}}^{(n)}(x)/n!$, $V_{\text{LJ}}^{(n)}$ is the $n$th derivative of the Lennard-Jones potential, $F_{\text{LJ}}(x)=-\lambda_1 (x)$ is the Lennard-Jones force, $k_{\text{eff}}\equiv k+2\lambda_2(x)$ is the effective spring constant, and $V_0(x,\delta x)$ includes constants and high-order nonlinear contributions from surface atomic forces to the harmonic potential. The equilibrium position $x_c$ is determined by balancing the static force from the Lennard-Jones potential and the restoring cantilever spring force, implying that $kx_c+F_{\text{LJ}}(x)=0$. Therefore, the Hamiltonian in Eq. \ref{Eq. A1} becomes
\begin{align}
H\approx\frac{p^2}{2m_{\text{eff}}}+\frac{1}{2}k_{\text{eff}}\delta x^2+V_0(x,\delta x).
\label{Eq. AA1} 
\end{align}
In the non-contact regime and for small oscillation amplitudes, the nonlinear contributions of $V_0(x,\delta x)$ are a perturbation to the harmonic potential. With this choice, at the single-phonon level, we employ the first-order perturbation theory to find the energy eigenvalues by writing the quantized displacement amplitude as $\delta x = x_{\text{zpf}}(a^\dag+a)$, where $x_{\text{zpf}}$ denotes the zero-point motion and $a$,$a^\dag$ are oscillator ladder operators. Therefore, the energy spectrum of Eq.~\ref{Eq. AA1} becomes
\begin{align}
E_n\approx\hbar\omega_{\text{eff}}(n+\frac{1}{2})+\sum_{k\geq2}{x_{\text{zpf}}^{2k}\lambda_{2k}(x)\expval{(a^\dag+a)^{2k}}{n}},
\label{Eq. A2}
\end{align}
where the first term is the quantum harmonic oscillator energy with $n$ enumerating the Fock levels, $\omega_{\text{eff}}=\sqrt{(k+k_{\text{LJ}})/m_{\text{eff}}}$ where $k_{\text{LJ}}=|V_{\text{LJ}}^{\left(2\right)}|$ is the spring constant induced by the Lennard-Jones potential, and the sum incorporates non-vanishing even-order contributions of the Lennard-Jones expansion series. The first two terms of the sum are
\begin{gather}
\expval{(a^\dag+a)^{4}}{n}=6n^2+6n+3, \\
\expval{(a^\dag+a)^{6}}{n}=20n^3+30n^2+40n+15.
\end{gather}
With these expressions, the energy spectrum of the nanomechanical qubit in Eq. \ref{Eq. A2} to the sixth-order correction can be written as
\begin{align}
E_n\approx\alpha_0+\alpha_1n+\alpha_2n^2+\alpha_3n^3
    \label{Eq. A5}
\end{align}
where 
\begin{gather}
\alpha_0=15\lambda_6x_{\text{zpf}}^6+3\lambda_4x_{\text{zpf}}^4+\hbar\omega_{\text{eff}}/2+V_{\text{LJ}}\left(x\right),\\\label{Eq. A8}
\alpha_1=40\lambda_6x_{\text{zpf}}^6+6\lambda_4x_{\text{zpf}}^4+\hbar\omega_{\text{eff}},\\ 
\alpha_2=30\lambda_6x_{\text{zpf}}^6+6\lambda_4x_{\text{zpf}}^4,\\
\alpha_3=20\lambda_6x_{\text{zpf}}^6. 
\end{gather}
Including more terms in the Taylor series, the energy converges to $E_n\approx\sum_{j\geq0}{\alpha_jn^j}$. For our purposes, the nonlinearity is mainly dominated by the first term in the summation of Eq. \ref{Eq. A2}. Supplemented by including the second term for higher accuracy, these two terms sufficiently approximate the qubit energy. Using the definition of the single-phonon relative anharmonicity, $\eta_r=(\omega_{21}-\omega_{10})/\omega_{10}$ where $\omega_{ij}=(E_i-E_j)/\hbar$, and Eq. \ref{Eq. A5}, we find
\begin{align}
 \eta_r\approx\frac{1+2r_1(x)}{1+r_1(x)+r_0(x)},
 \label{Eq. A10}
\end{align}
where $r_0(x)=2\hbar\omega_{\text{eff}}/(x_{\text{zpf}}^4V_{\text{LJ}}^{(4)}(x))$ and $r_1(x)=V_{\text{LJ}}^{(6)}(x)x_{\text{zpf}}^2/(4V_{\text{LJ}}^{(4)}(x))$. Aside from $V_{\text{LJ}}^{(n)}(x)$, which can be calculated directly from the Lennard-Jones potential, the nanomechanical qubit relative anharmonicity depends on $x_\text{zpf}=\sqrt{\hbar/(2m_{\text{eff}}\omega_\text{\text{eff}})}$. The unknown parameters in the $x_\text{zpf}$ expression are the native spring constant $k$ and the cantilever effective mass $m_{\text{eff}}$. These two quantities are given by $k = 3EI/L^3$ and $m_{\text{eff}}=k/\omega_c^2=0.2427\rho Lwt$, where $E$ is Young's modulus, $I = tw^3/12$ is the moment of inertia for the cantilever lateral mode, $\rho$ is the volume mass density, $w$ is the width, $t$ denotes thickness, $L$ is the cantilever thickness, and $\omega_c=1.015\sqrt{Ew^2/(\rho L^4)}$ is the isolated cantilever angular frequency \cite{giessibl2003advances,chen2021introduction,microtransducer}. Using the definitions used in Eq. \ref{Eq. A10}, one can also express the  nanomechanical qubit absolute anharmonicity ($\eta=\omega_{10}\eta_r$) as
\begin{align}
    \eta\approx\frac{x_{\text{zpf}}^4V_{\text{LJ}}^{(4)}(x)}{2\hbar}(1+2r_1(x)).
    \label{Eq. A11}
\end{align}
Eq. \ref{Eq. A11} suggests that $\eta$ is maximized at $x = x_0\approx1.24\sigma$, the point where $k_{\text{LJ}} = 0$. This is due to the dominance of $x_{\text{zpf}}^4$ over $V_{\text{LJ}}^{(4)}(x)$ for our given parameters. Since $x_{\text{zpf}}\propto\omega_{\text{eff}}^{-1/2}$, we have $\eta\propto\omega_{\text{eff}}^{-2}$, implying that $\eta$ is maximized at $x_0\approx1.24\sigma$, where $\omega_{\text{eff}}$ is at its minimum and equal to the isolated cantilever frequency ($\omega_{\text{eff}}\approx\omega_c$).

\section{III. Effective external coupling}
\label{C}
\begin{figure}
    \centering
    \includegraphics[width=1\linewidth]{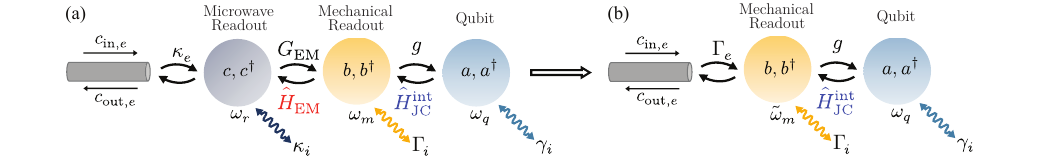}
    \caption{\textbf{Effective circuit quantum acoustodynamics setup.} \textbf{(a)} Interactions between coupled modes in the cQAD setup, described in Fig. 3(a) of the main text, can be reduced to \textbf{(b)} the Jaynes-Cummings interaction of circuit quantum electrodynamics between the qubit and mechanical readout resonator via the adiabatic elimination of the microwave mode $c$. $\gamma_i$, $\Gamma_i$, and $\kappa_i$ are the respective subsystem damping rates, ${\widetilde{\omega}}_m$ is the effective mechanical readout frequency, and $\Gamma_e$ is the effective external coupling rate.}
    \label{SI3 Fig}
\end{figure}

In this section, we show the circuit quantum acoustodynamics (cQAD) equations of motion and the effective external coupling rates of the system to microwave fields. We demonstrate that our system follows the already-established framework of circuit quantum electrodynamics (cQED). Consequently, one can extend quantum information processing methods with cQED to the cQAD setup. We begin with the Hamiltonian ${\hat{H}}_{\text{cQAD}}={\hat{H}}_0+{\hat{H}}_{\text{JC}}^{\text{int}}+{\hat{H}}_{\text{EM}}$ with
\begin{gather}
\frac{{\hat{H}}_0}{\hbar}\approx(1-\frac{\eta_r}{2})\omega_qa^\dag a+\frac{\eta_r\omega_q}{2}(a^\dag)^2a^2+\omega_mb^\dag b+\omega_rc^\dag c,\\
{\hat{H}}_{\text{JC}}^{\text{int}}=\hbar g(a^\dag b+a b^\dag),\\
{\hat{H}}_{\text{EM}}=\hbar g_{\text{EM}} c^\dag c(b^\dag+b), \label{Eq. B3}
\end{gather}
where the subsystem operators and coupling rates are labeled in Fig.~\ref{SI3 Fig}(a), and we accounted for the rotating wave approximation in the Jaynes-Cummings coupling. Eq.~\ref{Eq. B3} denotes the electromechanical coupling between the mechanical and microwave modes  \cite{aspelmeyer2014cavity,fink2016quantum,kalaee2019quantum}.
Here, the microwave resonance frequency modulation by the mechanical amplitude leads to vacuum electromechanical coupling strength $g_{\text{EM}}=X_{\text{zpf}}\partial\omega_r/\partial x$, where $X_{\text{zpf}}$ is the zero-point motion of the mechanical readout resonator, and $x$ denotes mechanical displacement \cite{aspelmeyer2014cavity,fink2016quantum}. The vacuum electromechanical coupling rate is $g_{\text{EM}}=q\omega_r X_{\text{zpf}}/2d$, where $q\equiv C_m/C_{\text{tot}}$ is the participation ratio, $C_m$ is the capacitance on the mechanical resonator, $C_{\text{tot}}$ is the total capacitance, and $d$ is the vacuum gap distance \cite{fink2016quantum,kalaee2019quantum} (see Fig.~3 in the main text and Fig.~\ref{SI2 Fig}(a)). Driving the microwave field with a tone at frequency $\omega_d$, the cQAD Hamiltonian becomes
\begin{align}
\frac{{\hat{H}}_{\text{cQAD}}}{\hbar}\approx{\omega}_qa^\dag a+\omega_mb^\dag b+\Delta_rc^\dag c+g(a^\dag b+ab^\dag)+G_{\text{EM}}(b^\dag c+bc^\dag),
\label{Eq. B4}
\end{align}
where $\Delta_r=\omega_r-\omega_d$,  $G_{\text{EM}}=g_{\text{EM}}\sqrt{n_d}$, and $n_d$ is the number of photons in the microwave resonator. We approximated the qubit as an oscillator and assumed the rotating wave approximation within the resolved-sideband cooling regime ($\kappa/\omega_m\ll1$) \cite{mirhosseini2020superconducting,aspelmeyer2014cavity}. Employing quantum input-output formalism \cite{gardiner1985input,meystre2021quantum} and Eq. \ref{Eq. B4}, the system equations of motion would be
\begin{gather}
\dot{a}=-(i\omega_q+\frac{\gamma_i}{2})a-igb-\sqrt{\gamma_i}a_{\text{in}}\label{Eq. B5},\\
\dot{b}=-(i\omega_m+\frac{\Gamma_i}{2})b-iga-iG_{\text{EM}}c-\sqrt{\Gamma_i}b_{\text{in}},\label{Eq. B6}\\ 
\dot{c}=-(i\Delta_{r}+\frac{\kappa}{2})c-iG_{\text{EM}}b-\sqrt{\kappa_i}c_{\text{in}}-\sqrt{\kappa_e}c_{\text{in},e}. 
\label{Eq. B7}
\end{gather}
where $\gamma_i$, $\Gamma_i$, and $\kappa=\kappa_i+\kappa_e$ are the respective subsystem damping rates, $a_\text{in},b_\text{in},c_\text{in}$ are Langevin noise operators due to intrinsic decay, $c_\text{in,e}$ is the input microwave field. For $G_{\text{EM}}/\Delta_r\ll1$, we adiabatically eliminate the microwave mode $c$. Fourier transforming Eq. \ref{Eq. B7}, we obtain the following steady-state solution
\begin{align}
c=\frac{iG_{\text{EM}}b+\sqrt{\kappa_i}c_{\text{in}}+\sqrt{\kappa_e}c_{\text{in},e}}{i\delta-\kappa/2},
\label{Eq. B8}
\end{align}
where $\delta=\omega_m-\Delta_r$. Inserting Eq. \ref{Eq. B8} into the Fourier transform of Eq. \ref{Eq. B6}, rearranging, and taking the inverse Fourier transform, we have
\begin{align}
\dot{b}=-\left(i{\widetilde{\omega}}_m+\frac{\Gamma}{2}\right)b-iga-\sqrt{\Gamma_i}b_{\text{in}}-\alpha\sqrt{\kappa_i}c_{\text{in}}-\alpha\sqrt{\kappa_e}c_{\text{in},e},
\label{Eq. B9}
\end{align} 
where $\alpha=iG_{\text{EM}}/(i\delta-\kappa/2)$, with the following effective parameters
\begin{gather}
{\widetilde{\omega}}_m\equiv\omega_m+\frac{4G_{\text{EM}}^2\delta}{4\delta^2+\kappa^2},\\
\Gamma\equiv\Gamma_i+\Gamma_e=\Gamma_i+\frac{4G_{\text{EM}}^2\kappa}{4\delta^2+\kappa^2},\label{Eq. B11}
\end{gather}
where $\Gamma_e=|\alpha|^2\kappa$ physically corresponds to the engineered Purcell decay rate of the mechanical readout resonator to the microwave transmission line as shown in Fig.~\ref{SI3 Fig}(b). This engineered coupling to the microwave lines allows the system to be controlled and read out using microwave fields. 

Incorporating effective parameters, the cQAD Hamiltonian is reduced to an effective Jaynes-Cummings Hamiltonian (Fig.~\ref{SI3 Fig}(b)) with engineered external coupling to the microwave fields. This system, including the weakly nonlinear nature of the qubit, is identical to cQED with transmon qubits (c.f., \cite{meystre2021quantum,blais2007quantum}). Similarly, the dispersive shift, $\chi=-g^2\eta/\Delta(\Delta+\eta)$ \cite{koch2007charge}, and the bus-mediated qubit-qubit coupling strength, $J=g_1g_2(1/\Delta_1+1/\Delta_2)/2$ \cite{majer2007coupling}, follow the results for transmon qubits, where $\Delta$ denotes qubit-cavity detuning, $g$ is the qubit-cavity coupling strength, and the subscript labels qubits.

\bibliography{main}